# Imaging the coexistence of superconductivity and a charge density modulation in $K_{0.73}Fe_{1.67}Se_2$ superconductor

Peng Cai[1], Cun Ye[1], Wei Ruan[1], Xiaodong Zhou[1], Aifeng Wang[2], Meng Zhang[2], Xianhui Chen[2], and Yayu Wang[1,†]

[1]*State Key Laboratory of Low Dimensional Quantum Physics, Department of Physics, Tsinghua University, Haidian, Beijing 100084, P.R. China*

[2]*Hefei National Laboratory for Physical Science at Microscale and Department of Physics, University of Science and Technology of China, Hefei, Anhui 230026, P.R. China*

We report scanning tunneling microscopy studies of the local structural and electronic properties of the iron selenide superconductor $K_{0.73}Fe_{1.67}Se_2$ with $T_C$ = 32K. On the atomically resolved FeSe surface, we observe well-defined superconducting gap and the microscopic coexistence of a charge density modulation with $\sqrt{2}\times\sqrt{2}$ periodicity with respect to the original Se lattice. We propose that a possible origin of the pattern is the electronic superstructure caused by the block antiferromagnetic ordering of the iron moments. The widely expected iron vacancy ordering is not observed, indicating that it is not a necessary ingredient for superconductivity in the intercalated iron selenides.

The iron-based superconductors have attracted tremendous recent interest due to the promise of solving high $T_C$ superconductivity (SC) via a new route [1-2]. Similar to the cuprates, the iron compounds also possess a complex phase diagram involving a variety of structural, electronic, and magnetic phases [3]. A key task in finding the mechanism of SC is to clarify the nature of each phase and its implications to SC. For example, the existence of antiferromagnetic (AF) ordering in close proximity to SC in the iron pnictides suggests that spin fluctuation may play a crucial role in mediating the Cooper pairing [4]. On the other hand, the discovery of nematic electronic ordering in the parent state of 122 pnictide suggests that stripe-like models may be highly relevant [5].

The intricate interplay between various phases is prominently exemplified in the new iron selenide superconductors $A_x Fe_y Se_2$ ($A$ = K, Rb, Cs, Tl, etc.) with $T_C$ above 30K [6-11]. Due to the distinct Fermi surface (FS) topology from their pnictide counterparts [12-14], $A_x Fe_y Se_2$ are expected to be ideal test grounds for theoretical models of iron-based SC. However, due to the chemical off-stoichiometry and spatial inhomogeneities in $A_x Fe_y Se_2$, a consensus regarding the lattice, electronic and magnetic structures of the SC phase is still lacking. Muon spin rotation (μSR) experiment on $Cs_{0.8}(FeSe_{0.98})_2$ detects the microscopic coexistence of SC and AF [15] and suggests that the two orders are closely related. Neutron scattering on $K_{0.8}Fe_{1.6}Se_2$ reveals the formation of Fe vacancy ordering and block checkerboard AF phase at elevated temperatures [16-17], which has prompted the proposal of Fe vacancy ordered Mott insulator as the parent state [18-19]. Transport and magnetization studies suggest that $A_x Fe_y Se_2$ is an AF insulator for Fe content $1.5 < y < 1.6$ and SC emerges only for $y > 1.8$ when many of the Fe vacancies are filled [9, 11, 20]. To make it more complicated, x-ray diffraction (XRD) [21] and transmission electron microscopy (TEM) [22] reveal the separation of $A_x Fe_y Se_2$ into multiple phases with varied $A$ and Fe contents.

Scanning tunneling microscopy (STM) represents an ideal probe for resolving these controversies owing to the unique capability of detecting the local structural and electronic properties simultaneously. In this letter we report atomic scale STM studies of the $K_{0.73}Fe_{1.67}Se_2$ single crystals with $T_C$ = 32K. On atomically resolved FeSe layer without Fe vacancy, we observe an unexpected $\sqrt{2} \times \sqrt{2}$ charge density modulation coexisting with the SC phase. We propose that the electronic superstructure is possibly caused by the block AF ordering in the Fe layer, which implies a microscopic coexistence of SC and AF.

High quality $K_{0.73}Fe_{1.67}Se_2$ single crystals are grown by the Bridgeman method [7]. The chemical composition is calibrated by using the energy-dispersive x-ray spectrometer mounted on the field emission scanning electron microscope (Sirion200). Fig. 1a shows the sharp SC transition at $T_C$ = 32K, and Fig. 1b reveals two consecutive phase transitions at much higher $T$. The increase of resistivity at $T_S$ = 546K is usually ascribed to the $\sqrt{5} \times \sqrt{5}$ Fe vacancy ordering, and the drop of susceptibility at $T_N$ = 535K indicates the appearance of AF ordering. For STM experiments, $K_{0.73}Fe_{1.67}Se_2$ single crystals are cleaved *in situ* at 77K and then immediately transferred into the STM stage, which stays at 5K in ultrahigh vacuum. All the data shown here are obtained at the base temperature $T$ = 5K. The STM topography is taken in the constant current mode, and the *dI/dV* spectroscopy is collected by standard ac lock-in method with a modulation frequency $f$ = 423Hz.

Figure 2(a) shows the schematic 122-type crystal structure of $KFe_2Se_2$, which consists of FeSe layers of edge-shared $FeSe_4$ tetrahetra separated by the intercalated K atoms. The crystal is expected to cleave between two adjacent FeSe layers, exposing the Se terminated surface decorated with the remnant K atoms (on average half of the K atoms are left on the surface). Fig. 2(b) displays a typical topographic image of cleaved $K_{0.73}Fe_{1.67}Se_2$ taken with sample bias $V_s = 0.4V$ and tunneling current $I_t = 5pA$. The surface shows a rather complex morphology consistent with our expectation. On top of the large flat terrace there are bright spots either as isolated spheres or small clusters and patches. Fig. 2(c) presents a line profile along the red broken line in Fig. 2(b) and shows that the height of the bright spots is ~ 1Å, consistent with the size of K atom. Fig. 2(d) displays the derivative plot of the topography of an area with rather sparse K atoms (marked by dotted green square in Fig. 2(b)), which uncovers an atomically clear square lattice. The lattice constant is around $a = 4Å$, in agreement with the distance between two adjacent Se atoms of the FeSe layer.

We note that the chance of seeing the atomically resolved surface shown above is about 10% among more than 100 approaches on cleaved $K_{0.73}Fe_{1.67}Se_2$. On the rest of the surfaces we cannot obtain high quality STM images at $T = 5K$. A possible explanation is that the majority phase of the sample is insulating so that high resolution STM imaging is difficult at low $T$. The atomically resolved FeSe surface reported here may represent the minority phase of the crystal. This is consistent with the recent finding of a phase separation into $\sqrt{5} \times \sqrt{5}$ Fe vacancy ordered insulating phase and vacancy-free SC phase on (110)-oriented $K_xFe_ySe_2$ film grown by molecular beam epitaxy (MBE) [23].

Figure 3(a) shows the spatially averaged $dI/dV$ curve taken directly on flat Se surface without K atoms. The $dI/dV(V_s, \boldsymbol{r})$ is approximately proportional to the local electron density of state (DOS) of the sample at location $\boldsymbol{r}$ with energy $\varepsilon = eV_s$ (the Fermi level $E_F$ lies at $V_s = 0$). On the negative bias side of Fig. 3(a), which corresponds to the occupied states, the DOS shows a rapid increase starting from about $V_s = -50$mV and reaches a peak near $V_s = -330$mV. Very similar features have been seen by angle-resolved photoemission spectroscopy (ARPES) on $A_x$Fe$_y$Se$_2$ and are ascribed to the appearance of hole-like bands below $E_F$ [24]. The positive bias spectrum corresponds to the unoccupied states that cannot be reached by ARPES. The $dI/dV$ curve also shows a similar increase of DOS from $V_s \sim$ 50mV till 300mV, although the features are not as sharp as that in the negative side. The overall $dI/dV$ profile therefore represents a broad gap-like DOS suppression between $\varepsilon = \pm 50$meV around the $E_F$.

Figure 3(b) displays the low bias $dI/dV$ curve with higher energy resolution. The most pronounced feature here is the superconducting gap located at $E_F$. The gap amplitude determined by the distance between the two shoulders [Fig. 3(b) inset] is $2\Delta \sim 14$meV, slightly smaller than that observed by ARPES on similar K$_x$Fe$_y$Se$_2$ crystals [12]. The ratio of $2\Delta/k_B T_C$ is around 5, which is beyond the BCS weak coupling limit but in the same range as the 122 iron pnictides [25]. A closer examination shows that the gap is incomplete with finite DOS at $E_F$ that is too large to be explained by thermal activation at $T = 5$K. The exact origin of the residual DOS is unclear at the moment. Another fine structure revealed by Fig. 3(b) is the two successive DOS jumps at $\varepsilon = -50$meV and $-66$meV, which agrees remarkably well with the ARPES spectrum showing the onsets of two hole-like bands at these energies [24]. The increase of DOS at $V_s = 50$mV, which is nearly symmetric to that at the negative bias, also becomes more pronounced in this energy scale.

Figure 4 displays high resolution close-up images on a small K-free FeSe surface, which reveals a highly unexpected feature. The bright spots marked by circles indicate the existence of a new charge density modulation pattern. The superstructure has a $\sqrt{2}\times\sqrt{2}$ unit cell with respect to the Se-Se square lattice. In general the feature is quite weak, and the maximum apparent height difference between the bright and dark lattice sites is merely 0.03Å. We argue that this superstructure is electronic in origin rather than from a surface reconstruction because the feature is strongly bias dependent. The superstructures are most pronounced at $V_s$ = -20mV and -50mV, and become weaker at other biases. The patterns cannot be clearly resolved for sample biases higher than ±100mV. The superconducting spectrum shown in Fig. 3(b) can be observed in any location of the superstructure, indicating the homogeneous coexistence of SC with the charge density modulation.

From the STM images alone we cannot exclude the possibility that the new electronic superstructure is a peculiar surface effect. Recently, the same superstructure has been found by XRD and TEM in similar $K_xFe_ySe_2$ crystals [21, 26]. In both experiments, the $\sqrt{5}\times\sqrt{5}$ Fe vacancy ordering is observed at high $T$, but a weak $\sqrt{2}\times\sqrt{2}$ superstructure emerges at low $T$. Although the two experiments give different pictures regarding the evolution of the two phases, they both indicate that the ground state of the SC phase has a $\sqrt{2}\times\sqrt{2}$ superstructure. The agreement between the bulk structural studies and our STM images suggests that the $\sqrt{2}\times\sqrt{2}$ superstructure is an intrinsic bulk property of the SC phase of $K_xFe_ySe_2$.

The coexistence of electronic ordering and SC is a rather common phenomenon in low dimensional superconductors including the cuprates, layered chacolgenides, and organic compounds. The mechanism of the electronic ordering and its implications to SC, however, could be a very challenging problem. There usually exists two totally different scenarios, one involves the FS nesting of itinerant electrons and the other involves local ordering of certain degree of freedom. A notable example is the checkerboard electronic ordering found in the cuprates [27-28]. While ARPES suggests that the superstructure is due to the charge density wave (CDW) formed by FS nesting at the antinodal region [29], comparison with neutron scattering data suggests that it is the charge density modulation accompanying the spin density wave order [30]. Next we discuss the relevance of these two pictures to the $\sqrt{2}\times\sqrt{2}$ superstructure observed here in $K_{0.73}Fe_{1.67}Se_2$.

The band structure of $K_xFe_ySe_2$ provides some hints for FS nesting induced CDW. Recent ARPES measurements in $K_{0.68}Fe_{1.79}Se_2$ [24] reveal a weak electron-like FS pocket (the β band) near the Γ point of the BZ with similar size and shape to that around the *M* point [Fig. 5(a)], implying a possible instability towards FS nesting between the two FS pockets. Moreover, the nesting vector between Γ and *M* will lead to a CDW order with $\sqrt{2}\times\sqrt{2}$ superstructure, making it an attractive proposal. However, the β band is very weak or even nonexistent in ARPES band maps [12], and LDA calculations cannot reproduce such a large electron-like FS pocket near Γ [13]. Both facts suggest that the β pocket may not be the cause of the nesting, but instead might be the consequence of folded FS from the *M* point when the $\sqrt{2}\times\sqrt{2}$ superstructure is formed.

Another more likely origin of the superstructure is certain symmetry breaking states,

especially the spin ordering in the Fe layer, which may cause charge density modulation on the Se layer. The magnetic structure of the $K_xFe_ySe_2$ system has been investigated by neutron scattering. In the $\sqrt{5}\times\sqrt{5}$ Fe vacancy ordered phase, a block checkerboard AF ordering is found with effective Fe moment as large as 3.3$\mu_B$ [16]. Recent theoretical calculations based on the local exchange model suggest that the block AF ordering is also a degenerate ground state of the vacancy-free phase with tetragonal symmetry [31]. Furthermore, due to the strong magneto-structural coupling the energy can be further reduced when the four Fe lattices forming the block contract slightly towards the center [31]. Interestingly, this contracted block AF spin ordering will exactly induce a $\sqrt{2}\times\sqrt{2}$ charge density modulation on the Se layer due to the different environment of the nearest neighbor Se sites, as illustrated in Fig. 5(b). If this is indeed the correct picture for the superstructure, then what we found is actually a microscopic coexistence of SC and AF in $K_xFe_ySe_2$, consistent with the μSR result [15]. However, we emphasize that at the current stage this scenario is still a speculation because the magnetic structure of the vacancy-free $K_xFe_2Se_2$ compound is unknown due to the lack of pure phase samples. Moreover, we cannot exclude other possible symmetry breaking state, such as Fe orbital ordering, as the likely source of the superstructure.

Another important finding of this work is the absence of Fe vacancy in the SC phase of $K_{0.73}Fe_{1.67}Se_2$. Although the Fe layer lies at 2Å below the surface, the defect on Fe site is expected to create pronounced electronic states that can be detected by STM. In the $FeSe_{1-x}Te_x$ and LaOFeAs compounds [32-33], the Fe site defects can be clearly resolved in STM images as bias-dependent dumb-bells due to the tetrahedral bonding between Fe and the surface Se/As atoms. However, our high resolution STM images on atomically flat FeSe surface of $K_{0.73}Fe_{1.67}Se_2$ [Fig. 4] show no sign of Fe vacancy despite extensive search. Recent STM studies on (110)-oriented MBE grown $K_xFe_ySe_2$ films also find that the SC phase is the stoichiometric $K_xFe_2Se_2$ without Fe vacancy. The two complimentary STM results demonstrate unambiguously that Fe vacancy ordering is not an essential ingredient for SC in $K_xFe_ySe_2$. This observation calls for a reexamination of the basic assumptions regarding the mechanism of SC. Unlike the viewpoint of doping the iron vacancy ordered AF Mott insulator, our STM results suggest that the SC phase of $A_xFe_ySe_2$ may be simply viewed as perfect FeSe layers intercalated by K atoms. Recall that $T_C$ of the 11-type FeSe SC can be enhanced from 8.5K [34] to near 37K by high pressure [35]. We may hypothesize that in $A_xFe_ySe_2$ the $A$ intercalations provide the charge doping and chemical pressure necessary for the stoichiometric FeSe layers to have high $T_C$.

In summary, atomic scale STM studies on $K_{0.73}Fe_{1.67}Se_2$ single crystals provide new clues regarding the phase diagram of the intercalated iron selenides. The vacancy-free FeSe layer shows the microscopic coexistence of SC and a charge density modulation, which may be caused by the block AF ordering in the Fe layer. The Fe vacancy ordering is not essential for SC but instead may be related to the insulating phase that is spatially separated from SC. The growth and studies of pure phase superconducting and insulating $K_xFe_ySe_2$ crystals will help to further clarify these issues.

We thank Xi Chen, Donglai Feng, Zheng-Yu Weng, Tao Xiang, Guang-Ming Zhang, Xingjiang Zhou, and especially Jiangping Hu for helpful discussions.

[†] Electronic address: yayuwang@tsinghua.edu.cn

Figure Captions:

FIG 1. (a) The Meissner effect ($H_\perp = 10$Oe) and resistivity curves show sharp SC transition at $T_C = 32$K. (b) The increase of resistivity at $T_S = 546$K is due to the iron vacancy ordering, and the drop of magnetization ($H_{//} = 50$kOe) at $T_N = 535$K indicates the AF ordering.

FIG 2. (a) The schematic crystal structure of $KFe_2Se_2$. (b) STM topography of cleaved $K_{0.73}Fe_{1.67}Se_2$ obtained at sample bias $V_s = 0.4$V and tunneling current $I_t = 5$pA. (c) Line profile along the red broken line in (b) shows that the height of the bright spots is ~ 1Å, consistent with the size of K atoms. (d) Derivative plot of the topography of the area marked in (b) uncovers an atomically clear square lattice with lattice constant $a$ ~ 4Å.

FIG 3. (a) Spatially averaged $dI/dV$ spectrum measured on the flat FeSe surface without K atoms. (b) Spectrum zoomed into the low energy range shows a well-defined SC gap with amplitude $2\Delta$ ~ 14meV (inset). The arrows indicate the gap-like features at ±50mV.

FIG 4. High resolution images on the FeSe surface taken with different sample biases ($V_s = \pm10$mV, $\pm20$mV, $\pm50$mV) reveal a bias dependent $\sqrt{2}\times\sqrt{2}$ superstructure.

FIG 5. (a) Schematic ARPES-measured FS (black) shows three electron-like FS sheets, two around $\Gamma$ and one around $M$. The red arrow along the $(\pi, \pi)$ direction indicates the possible nesting vector between the $\beta$ and $\gamma$ FS pockets. The red broken line shows the reconstructed BZ by the $\sqrt{2}\times\sqrt{2}$ superstructure. (b) Schematic structure of the top Se surface and the underneath Fe layer with the proposed block AF structure. The four Fe lattices forming a ferromagnetic block (blue solid squares) tend to contract slightly toward the block center to further reduce the energy. The Se lattice will then form a $\sqrt{2}\times\sqrt{2}$ superstructure (red broken square) because the NN Se atoms become inequivalent.

**References:**

[1] Y. Kamihara *et al.*, J. Am. Chem. Soc. **130**, 3296 (2008).

[2] X. H. Chen *et al.*, Nature **453**, 761 (2008).

[3] K. Ishida, Y. Nakai, and H. Hosono, J. Phys. Soc. Jpn. **78**, 062001 (2009).

[4] C. de la Cruz *et al.*, Nature **453**, 899 (2008).

[5] T. M. Chuang *et al.*, Science **327**, 181 (2010).

[6] J. Guo *et al.*, Phys. Rev. B **82**, 180520 (2010).

[7] A. F. Wang *et al.*, Phys. Rev. B **83**, 060512 (2011).

[8] Y. Mizuguchi *et al.*, Appl. Phys. Lett. **98**, 042511 (2011).

[9] M.-H. Fang *et al.*, Europhys. Lett. **94**, 27009 (2011).

[10] R. H. Liu *et al.*, Europhys. Lett. **94**, 27008 (2011).

[11] D. M. Wang *et al.*, Phys. Rev. B **83**, 132502 (2011).

[12] Y. Zhang *et al.*, Nature Mater. **10**, 273 (2011).

[13] T. Qian *et al.*, Phys. Rev. Lett. **106**, 187001 (2011).

[14] D. Mou *et al.*, Phys. Rev. Lett. **106**, 107001 (2011).

[15] Z. Shermadini *et al.*, Phys. Rev. Lett. **106**, 117602 (2011).

[16] W. Bao *et al.*, Chin. Phys. Lett. **28**, 086104 (2011).

[17] V. Y. Pomjakushin *et al.*, Phys. Rev. B **83**, 144410 (2011).

[18] R. Yu, J.-X. Zhu, and Q. Si, Phys. Rev. Lett. **106**, 186401 (2011).

[19] X.-W. Yan *et al.*, Phys. Rev. Lett. **106**, 087005 (2011).

[20] Y. J. Yan *et al.*, arXiv:1104.4941 (2011).

[21] A. Ricci *et al.*, Supercond. Sci. Tech. **24**, 082002 (2011).

[22] Z. Wang *et al.*, Phys. Rev. B **83**, 140505 (2011).

[23] W. Li *et al.*, arXiv:1108.0069 (2011).

[24] L. Zhao *et al.*, Phys. Rev. B **83**, 140508 (2011).

[25] Y. Yin *et al.*, Phys. Rev. Lett. **102**, 097002 (2009).

[26] J. Q. Li *et al.*, arXiv:1104.5340v1 (2011).

[27] J. E. Hoffman *et al.*, Science **297**, 1148 (2002).

[28] T. Hanaguri *et al.*, Nature **430**, 1001 (2004).

[29] K. M. Shen *et al.*, Science **307**, 901 (2005).

[30] S. Sachdev, and S.-C. Zhang, Science **295**, 452 (2002).

[31] J. Hu *et al.*, arXiv:1106.5169 (2011), and private communications with Jiangping Hu.

[32] T. Hanaguri *et al.*, Science **328**, 474 (2010).

[33] X. Zhou *et al.*, Phys. Rev. Lett. **106**, 087001 (2011).

[34] F. C. Hsu *et al.*, Proc. Natl. Acad. Sci. **105**, 14262 (2008).

[35] S. Medvedev *et al.*, Nature Mater. **8**, 630 (2009).

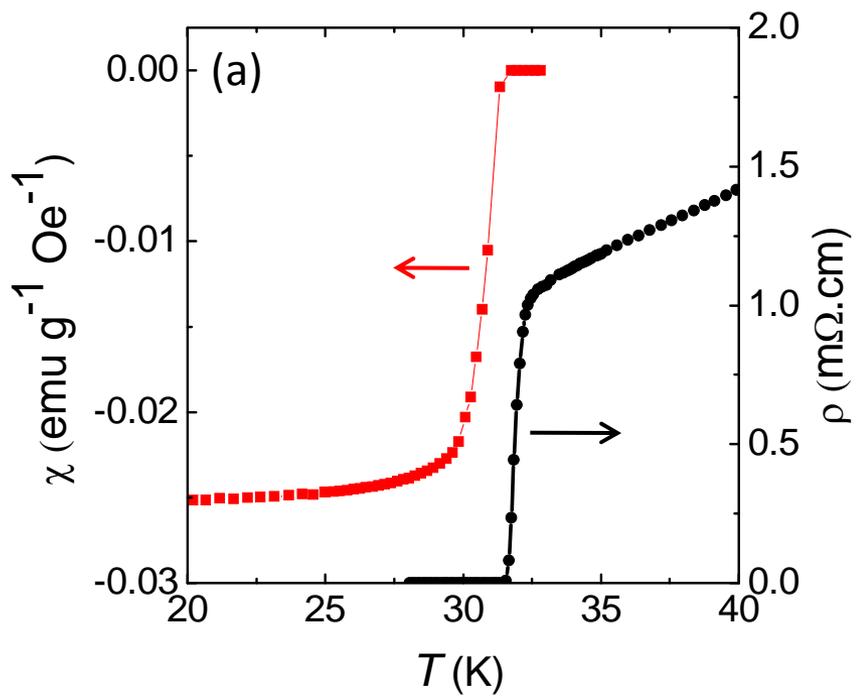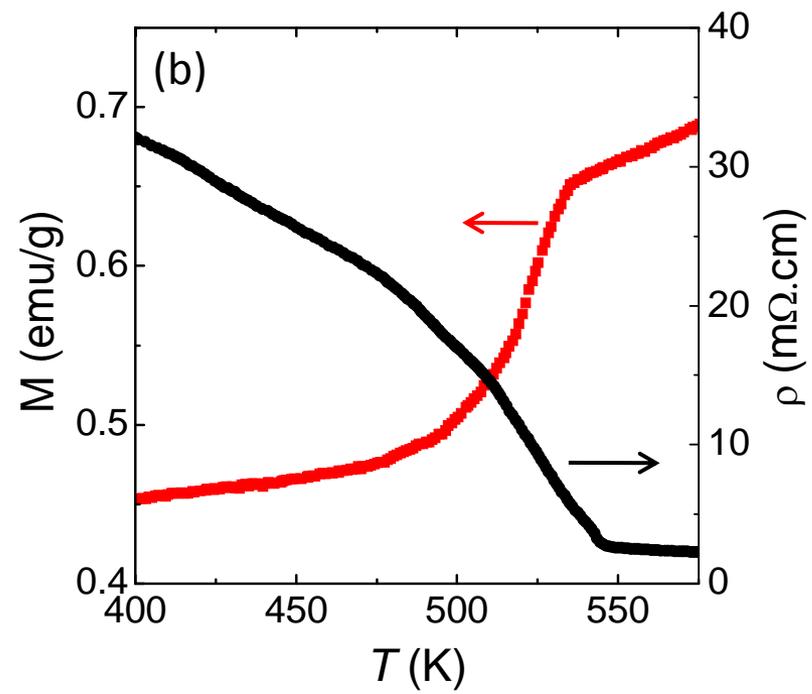

Figure 1

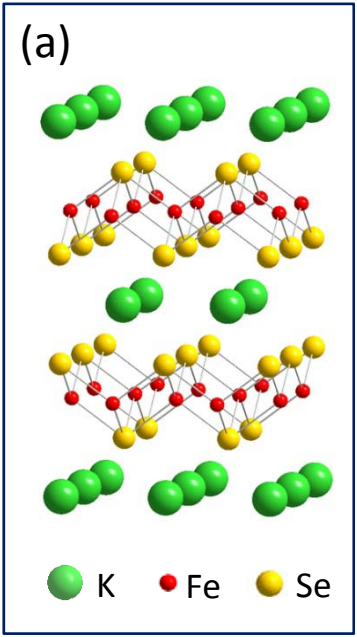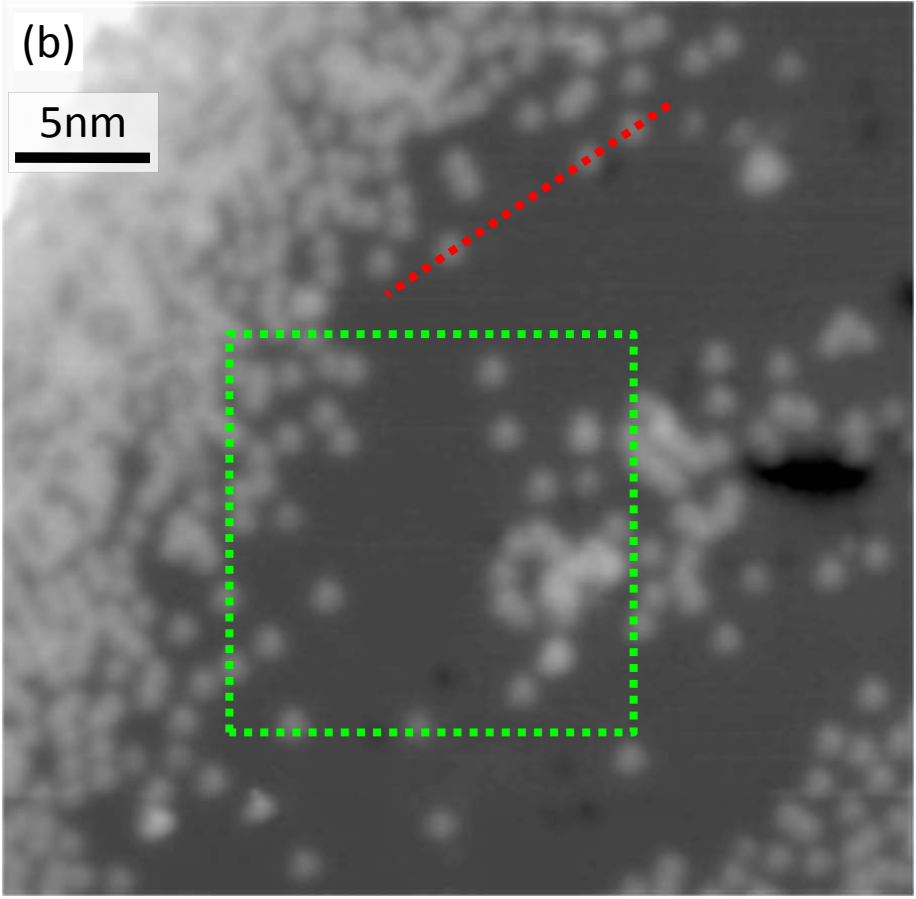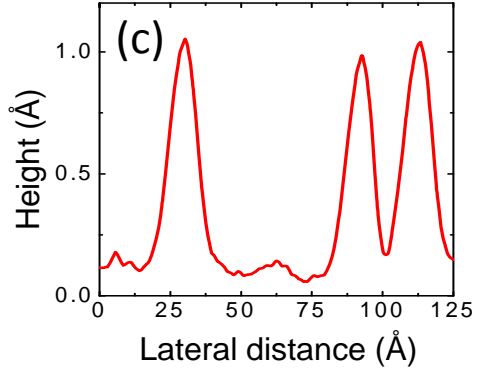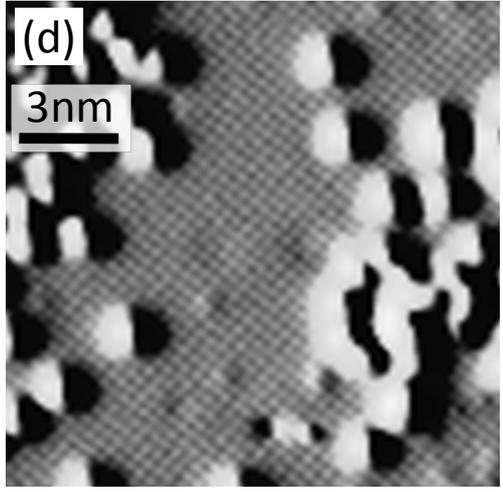

Figure 2

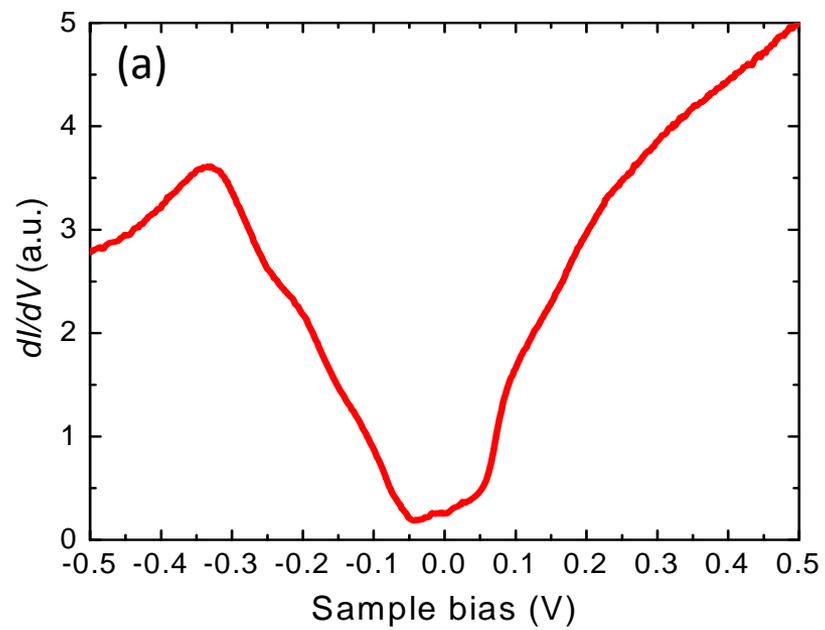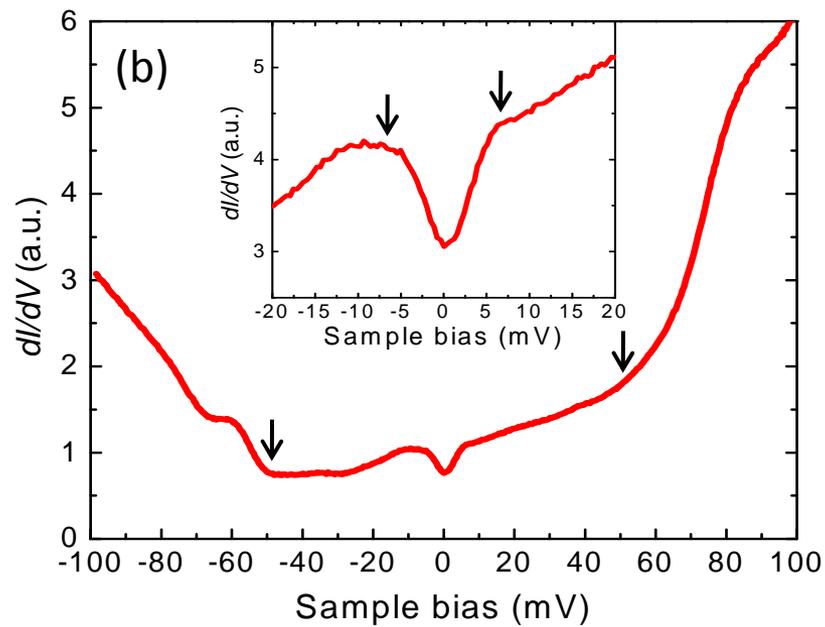

Figure 3

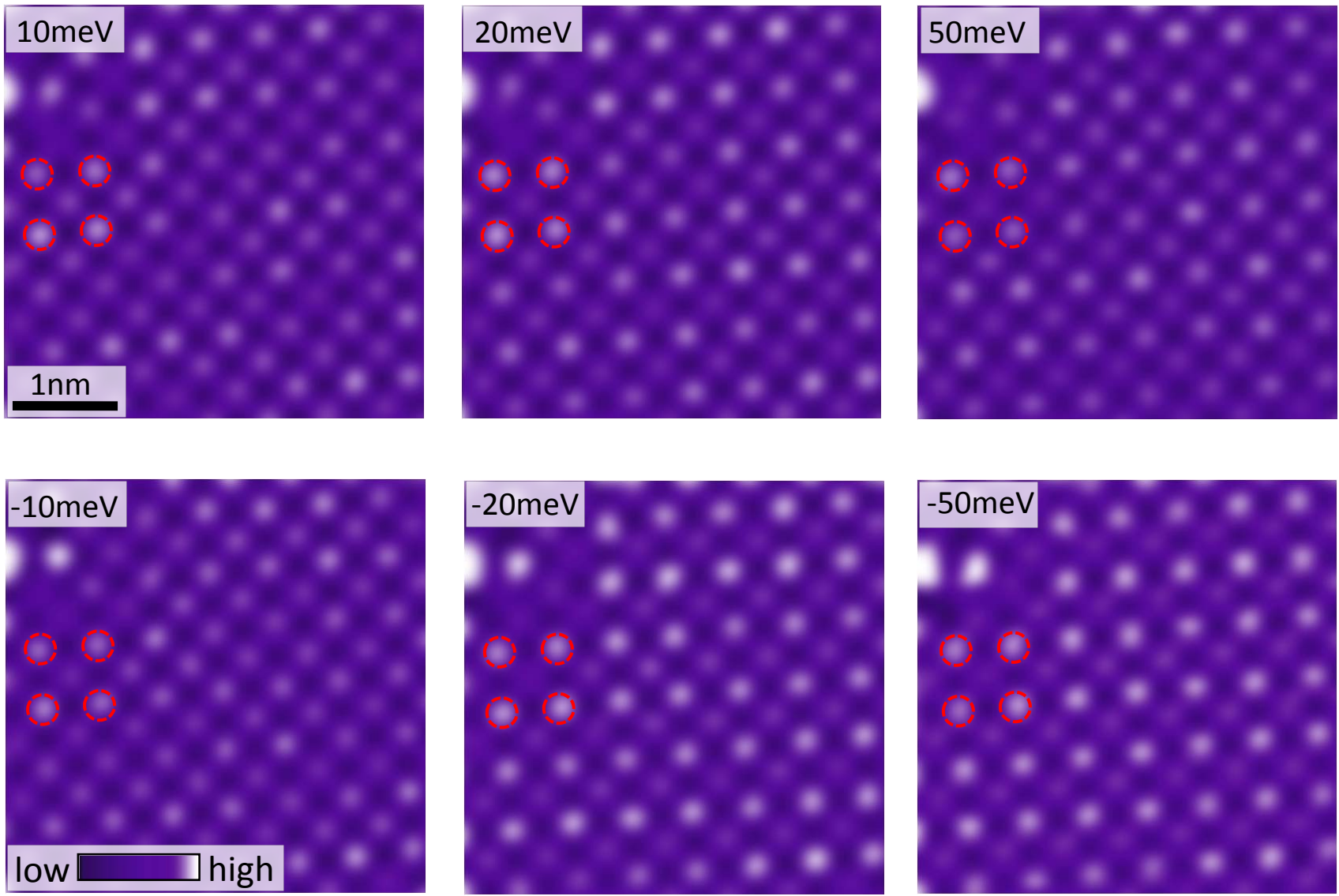

Figure 4

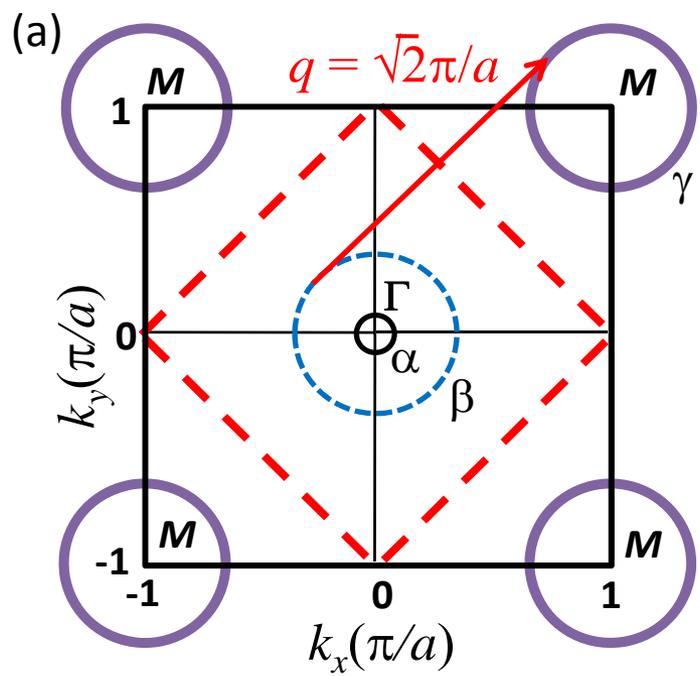 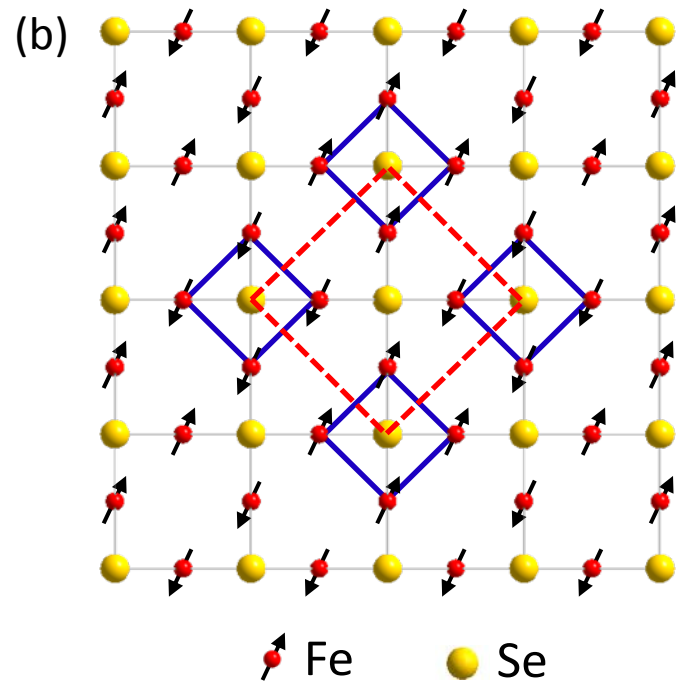

Figure 5